Publish-and-Flourish: decentralized co-creation and curation of scholarly content

Emilija Stojmenova Duh (1,2), Andrej Duh (1), Uroš Droftina (1),
Tim Kos (1,3), Urban Duh (1), Tanja Simonič Korošak (1,2), Dean Korošak *(1,4,5)

(1) Infinitcodex, Ltd., Maribor
(2) University of Ljubljana, Faculty of Electrical Engineering
(3) Institute of Mathematics, Physics and Mechanics, Ljubljana
(4) University of Maribor, Faculty of Medicine, Institute for Physiology
(5) University of Maribor, Faculty of Civil Engineering, Transportation Engineering and Architecture

Abstract

Scholarly communication is today immersed in publish-or-perish culture that propels noncooperative behavior in the sense of strategic games played by researchers. Here we introduce and describe a blockchain based platform for decentralized scholarly communication. The design of the platform rests on community driven publishing reviewing processes and implements cryptoeconomic incentives that promote cooperative user behavior. Key to achieve cooperation in blockchain based scholarly communication is to transform today's static research paper into a modifiable research paper under continuous peer review process. We describe and discuss the implementation of a modifiable research paper as a smart contract on the blockchain.

Keywords: scholarly communication, co-operation, co-creation, blockchain, decentralisation, smart contract, continuous peer review, publish-and-flourish

*Corresponding author: dean.korosak@um.si, University of Maribor

1.Introduction

The Internet and social media in particular are the enablers of media convergence, a phenomenon characterized with flow of content and migration of users, linking together content, communication and computation (Jenkins 2013; Duh 2016). But besides positive effects of convergence such as the rise of collective processes of information consumption and consumer-generated media, we are recently witnessing the emergence of fast traveling fake news influencing collective human decisions (Vosoughi 2018; Duh 2018), helped also by spreading of automated content generation (Lazer 2018). Even reporting on results of scientific discoveries in traditional academic journals is not immune to overhyped or even fake claims that cause irreproducibility and distrust (Kirchherr 2017).

At first, fast migration and automation of content, and trust seem orthogonal to each other. However, the application of blockchain technologies (Wust 2017) holds a promise to profoundly change and decentralize scholarly communication (Van 2017; Janowicz 2018), bringing together spreadability, migration credibility and trust, based on open science principles (Bartling 2014). Blockchains, in short, are digital ledgers of cryptographically validated transactions distributed in a network of nodes that validate each new block of transactions through some consensus mechanism before that block is immutably appended to the ledger (Nakamoto 2008; Buterin 2014; Tapscott 2016; Swan 2015). Blockchains are architecturally and politically decentralized: there is no single point of failure in the infrastructure and there is no single governing body of a blockchain. All nodes in the blockchain network must, however, with a consensus agree on one state of a blockchain, therefore making blockchain a logically centralized system (Buterin 2017).

Like blockchain, science is architecturally (research infrastructures do not have a single point of failure) and politically (research is not governed by a single scientific authority) decentralized (Bruecher 2018). Science is, however, logically centralized (scientific community must agree on one state -- scientific truth) through "a process that does lead to a broadly shared consensus. It is arguably the only social process that does." (Romer 2015). In science, this consensus is reached through scholarly communication.

Building of a truly decentralized and trusted scholarly communication model "prompts the articulation of the functions that academic publishing provides and how, if these are still required, they might be provided in decentralized models." (Swan 2015, p63).

Academic publishing (in printed and electronic form) is a prevailing medium of scholarly communication that in the current form suffers from inefficiencies including slow, incomplete, inaccurate and unmodifiable communication (Nosek 2012).

Today, there exist three models of academic publishing: paywalled (peer reviewed, free or fee for authors, readers pay to access the papers), open access (peer reviewed, fee for authors to publish, free access to published papers), and preprint repository (self-archiving or centralized archives of preprints, no fees, but no peer-review either).

Consolidation of academic publishing industry resulted in a market with limited competition and with ever increasing profits for the a few large players (Lariviere 2015; Bogich 2016; Schmitt 2015). Large profits are possible because the marginal costs of electronic publishing have no lower boundary (practically zero), and because paywalled or open access academic publishing stands on non-compensated work of researchers in three-fold roles: authors, reviewers and editors. Academic publishing was described as a coordination game (Bergstrom 2001) where users (authors, reviewers and editors) coordinate at journals and are stuck in one of the equilibria where they "continue to pay huge rents to owners of commercial journals" (Bergstrom 2001).

Large publishing companies have positioned themselves as centralized authorities, as gatekeepers of the quality of published research and providers of reputation for authors whose academic careers in turn depend on the products these same publishers provide. Academic life is immersed in publish-or-perish culture, the growing pressure on researchers to keep rapidly publishing scholarly content to survive in academia, resulting in fast-growing volume of published papers (Papatheodorou 2008; Ioannidis 2018). Such accelerated growth of published scholarly content slows down the progress in large scientific fields (Chu 2018) and affects science trustworthiness when the quantity of published work is used as dominant metric for evaluation of researchers (Grimes 2018). Trustworthiness in science is further eroded by arrival of predatory journals (Beall 2018) the existence of which "is only the most pathetic facet of a much deeper cultural problem within science – a form of comic relief on the backdrop of a tragedy, which we should all take as a reminder of how far from our goals we have strayed." (Amaral 2018).

At the core of scholarly communication is peer review, a recommendation process underlying publication decisions that should guard the trust and guarantee high quality of published work. However, under publish-or-perish culture, peer review became flawed, non-transparent, inconsistent and biased (Smith 2006; Hatton 2017; Lee 2013). The current situation in academic publishing market where the scientific journals operate without transparency of peer review process (Lee 2017; Taichman 2017; Tomkins 2017) is harmful to scientific progress (Ioannidis 2005; Buranyi 2017; Smaldino 2016), limits the participation of researchers from developing economies, is biased against young researchers, and creates wrong incentives for authors (Edwards 2017; Martinson 2017; Stephan 2017; Higginson 2016).

Are there ways out of the current scholarly communication paradigm? In terms of a coordination game, can the community of researchers move from current publish-or-perish equilibrium into publish-and-flourish equilibrium?

Suggestions to change the current paradigm include morphing the self-archiving preprint repositories into the arXivs of the future (Pepe 2017), using disruptive technologies such as peer-to-peer applications changing distributions of financial resources between researchers and institutions (Crous 2017), and decentralized blockchain solutions (Swan 2015; Janowicz 2018; Avital 2018; Van Rossum 2017) by creating platforms for open science (Bartling 2014) on the blockchain.

Following the analysis of inefficiencies of today's scholarly communication, a six-stage process approach of modifying current system into an open science platform was proposed (Nosek 2012) with the implementation of open, continuous peer review as the last stage of the transformation process. Key to the shift from publish-and-perish to publish-and-flourish academic culture is a profound change in incentives for authors: "...the scientist's ultimate objective is no longer to get published, because everything is published. The objective is to influence future ideas and investigations, that is, what should be the key incentive in the first place." (Nosek 2012).

Here, we suggest that for such a shift a radical change of the concept of research paper itself is necessary -- a change from unmodifiable to modifiable published scholarly content that by definition demands open and continuous review process. We show that distributed ledger technology (i.e. blockchain technology) based platform solution tailored to support open scholarly communication provides users (teachers, researchers, students) with mechanisms for building trust and reputation, protects privacy and operates with transparency and access through a decentralized peer-to-peer network. In particular, we propose a design for a platform that supports the implementation of encoding scholarly content into smart contracts -- a code that runs on a blockchain. Such encoding allows the content to function as a computable autonomous entity that can interact with other computable autonomous entities on a blockchain and with human users on the platform.

2. Theoretical motivations

Interesting set of experiments (Maxwell 2017) explored how blockchain technology as a new medium could change storytelling by considering stories as currency -- in this case as cryptocurrency, and noticed that both competitive and cooperative modes of user actions are induced by the distributed organisation of the blockchain. This opens up an intriguing possibility to explore the role of blockchain technology in strategic games (Fudenberg 1991) and in resolving social dilemmas occuring in scholarly communication set as game theoretical problems (Gall 2018; Leek 2011; Ellison 2002; Lacetera 2009; Kiri 2018; Gall 2017).

Let's consider two core processes in scholarly communication: publishing (making information available to public) and peer reviewing (recommendation process to publish/reject a paper) as games played by rational agents acting in both roles: as authors and as reviewers.

The quality of published works and therefore trustworthiness in science critically depends on the social norms of peer reviewing process (Ellison 2002). Diligent, responsible and honest peer reviews constitute a valuable resource of the scientific community -- the scientific commons (Stafford 2018). Under publish-or-perish pressure a researcher will prioritise her/his own publication output over providing quality reviews of other researchers' work, getting the benefit B > 0 (obtaining a review from the commons) without the effort (cost) e > 0 (contributing reviews to the commons). Each researcher can then play a strategy C (cooperate), contributing reviews with the pay-off B-e or can play a strategy D (defect) with the pay-off B if also enough other researchers (for instance more than N) in the community cooperate. If there are not enough cooperators in the community the pay-off of a cooperator is e, and the pay-off of a defector 0. The game structure is given in Table 1.

|   | #C > N | #C < N |
|---|--------|--------|
| C | B - e  | -e     |
| D | B      | 0      |

Table 1: The tragedy of the commons game structure.

With the pay-off structure: B > B-e > e > 0, a rational, strategic choice is to play D with the pay-off 0, thus exhausting the commons and causing "the tragedy of the commons" (Hardin 1968; Kuhn 2017) in this case the scientific commons.

Social mechanisms like indirect reciprocity (Nowak 1998) built on reputation of community members were shown to be able to resolve commons dilemma, sustain cooperation (and thus preserve the commons) in certain repeated games and help overcome the "tragedy of the commons" (Milinski 2002). More generally, for a successful resolution of social dilemma the interventions in strategies should focus on four key parts: information, identity, institutions and incentives (Van Vugt 2009).

The key principles of blockchain technology as enabler of decentralized scholarly communication are perfectly aligned with these foci of interventions and their core motives: understanding, belonging, trusting and self-enhancing (Van Vugt 2009). Blockchain, a distributed immutable database with equal and full transparency of information to all users without centralized authority, reduces uncertainty and increases understanding. Transactions based on pseudonymous peer-to-peer communication, with a choice for a user to opt for proof of identity to others, induces a strong sense of community. Integrity of the blockchain is protected with algorithms and rules defining consensus mechanisms that create decentralized trust and provide social and (crypto)economic incentives to reward honest work (proof-of-X) and increase reputation of users.

The ability of the blockchain - "the trust machine" (Economist 2015) - to act as a generator of decentralized trust without any central authority is in our view key to promote cooperation between users, to serve as a indirect reciprocity enabler and reputation builder with perfect memory accessible to anyone. Cooperation is possible even in games with defection (cheating) as dominant strategy in community of users which frequently interact such as prisoner's dilemma (Kuhn 2017; Fudenberg 1991; Ellison 1994). Discounting future payment in repeated games (Fudenberg 1991) can maintain cooperation if players are patient and value future earnings enough. Consider an example of a similar game between 2 players as before with benefit B = 2 and loss e = 1. A two-person variation is then given by the following structure shown in Table 2.

|   |   | B |   |
|---|---|---|---|
|   |   | C | D |
| A | C | 1,1 | -1,2 |
|   | D | 2,-1 | 0,0 |

Table 2: Prisoner's dilemma game payoff structure.

A repeated game in which each player adopts a strategy is played in consecutive rounds with payoff $u_{A,k}$ for player A in round k. A's average payoff in infinite game is:

$$<u_A> = (1-\delta)\sum_{k=0}^{\infty} \delta^k\, u_{A,k}, \qquad (1)$$

where δ = (0,1) is a discount factor -- a simple measure of player's patience for future payments and trust, or a measure of how the players value reputation (Friedman 2007). Figure 1 shows a "grim" strategy of player A who plays C (cooperation) as long as the other player cooperates. As soon as player B defects (plays D), the player A switches to D in the next and all following rounds.

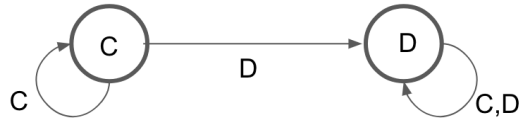

Figure 1: Grim strategy and transitions in repeated game as finite state machine

With such strategy and with the payoffs given in Table 1 the players will cooperate if average payoff for cooperation is greater than the average payoff for deviating (defecting):

$$<u_C> \geq <u_D>. \qquad (2)$$

Average payoff for cooperating is:

$$<u_C> = (1-\delta)(1 + \delta + \delta^2 + \ldots) = 1, \qquad (3)$$

while for defecting after player A defects in the first round we have:

$$<u_D> = (1-\delta)(1 + 0\delta + 0\delta^2 + \ldots) = 2(1-\delta). \qquad (4)$$

.

Therefore, a sustained cooperation requires sufficient patience with discount factor δ >= 1/2. In large populations of players with random matching maintaining cooperation requires larger values of discount factor (Ellison 1994) than in the two player games. If reputations (histories) of players in population of N + 1 players are known but cannot be communicated or publically shared in population then cooperation will exist when δ >= 1 - 1/2N (Friedman 2007). However, in situations where players start with and keep good reputation if they play (using a version of grim strategy as before) C against players with good reputations and play D against players with bad reputations, cooperation is maintained again with δ >= 1/2.

We assert that even in one-shot games describing situation in scholarly publishing as social dilemma, the blockchain could, as a generator of trust, act as a promoter of cooperation. Let two players (A, B) compete for priority in publishing a paper describing their research (Gall 2018). Each can either honestly report their research without additional effort e = 0 (choose to play C) or can spend an effort e > 0 to hype up the paper to make it appear more attractive (choose to play D). The expected payoff for researcher A depends on the reward for publishing R, the choice of effort and a function $P(e_A, e_B)$ describing the probability to publish that depends on the choices of both researchers:

$$u_A = P(e_A, e_B)R - e_B. \qquad (4)$$

We interpret $P(e_A, e_B)$ as the ability of scientific community as a collective to discern between honest research reporting and hyped up or even fraudulent papers. In publish-or-perish culture, papers that appear more novel and exciting tend to get publishing priority over honest reporting. This is modeled with $P(e_A = e, e_B = 0) = 1$ and $P(e_A = 0, e_B = e) = 0$ for player A, describing publication bias found in peer review practice (Lee 2013). When the effort of both players is the same, publishing probability equals a coin flip: $P(e_A = e, e_B = e) = 1/2$, reflecting difficulties in recognizing honest research reporting in current pre-publication review system with small number of reviewers involved in recommendation process. Choosing R = 4 and e = 1, we have the following structure for this game shown in Table 3.

|   | B |   |
|---|---|---|
|   | C | D |
| A C | 2,2 | 0,3 |
| A D | 3,0 | 1,1 |

Table 3: Example of prisoner's dilemma game capturing social dilemma in publish-or-perish academic culture.

This game portrays a social dilemma since even though mutual cooperation (C, C) offers better payoffs to players, mutual defection -- publish-or-perish behavior -- is the dominant strategy for both players, and therefore (D, D) is the game's equilibrium.

Blockchain based solutions for publish-or-perish induced problems in scholarly communication must include mechanisms and incentives that promote cooperation between researchers to, in the sense of "publication game", move the game equilibrium towards publish-and-flourish state (C, C). For instance, incentives for the community and mechanisms for lowering the publication bias so that $P(e_A = e, e_B = 0) = 1/2$ would lead, ceteris paribus, to the following game structure shown in Table 4.

|   | B |   |
|---|---|---|
|   | C | D |
| A C | 2,2 | 0,1 |
| A D | 1,0 | 1,1 |

Table 4: Example of a coordination game capturing social dilemma in publication game.

The game with the structure shown in Table 4 is an example of a coordination game (Cooper 1999) that has two pure equilibria: (C,C) and (D,D) and a mixed equilibrium where both players choose (D,D) with probability p = 0.5. Obviously, (C, C) publish-and-flourish equilibrium is preferable outcome for both players, but without communication between players prior to choosing strategy, coordination failures (i.e. choosing (D,D)) occur in this types of games (Cooper 1992). However, using mutual perfect communication prior to playing always leads to (C, C) equilibrium, while in games with noisy communication the selection of strategy C by both players depends on the communication signal (Carlsson 1993). In blockchain supported scholarly communication such signals are encoded in immutable memory of past events visible to all members of the community.

To promote and maintain users' cooperation and trustworthy publishing and reviewing blockchain supported processes, the key elements -- research papers and reviews -- of these processes must not be static objects. Research papers must, therefore, become modifiable entities under persistent review process. A research paper is, in a sense, a contract between author(s) and the community (depending on, for instance, strategic games rules as discussed before), so it is fitting to implement it as a smart contract (Szabo 1997) on a blockchain as discussed in the following section.

3. The I8X platform

In this section we introduce and describe INFINITCODEX (I8X) ([https://dev.infinitcodex.com](https://dev.infinitcodex.com)), scholarly publishing and reviewing blockchain based technology platform (Figure 2). The I8X platform enables and supports entwining the two core processes of scholarly communication -- publishing and reviewing -- by representing scholarly content as smart contracts and incorporating cryptoeconomical incentives for building reputation in a trustworthy community of users.

The two processes are embedded in the protocol and services layer of the I8X technology stack to support development of applications in the application layer. The infrastructure layer provides compute, storage, database and virtualization support, the protocol layer defines blockchain network participation requirement and rules, method and protocol of consensus, while the services layer contains blockchain services to enable development of applications and connections to other technologies.

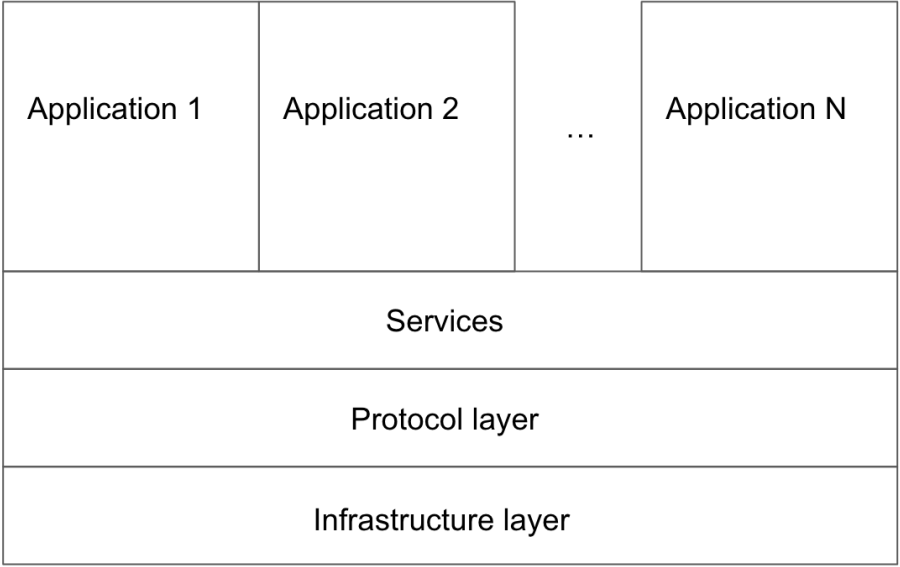

Figure 2: Technology stack of the I8X platform.

The two core processes were designed with three key principles in mind:
- transparency,
- integrity and
- engaging the members of the community.

Transparency and integrity are guarded by the protocol and network governance provided by the community members. I8X blockchain protocol and developed services such as smart contracts contain incentives (cryptoeconomic as well as social through reputation mechanisms) to stimulate community engagement in all platform activities performed by community members. The functional overview of the interdependence of processes, user interaction and community is shown in Figure 3.

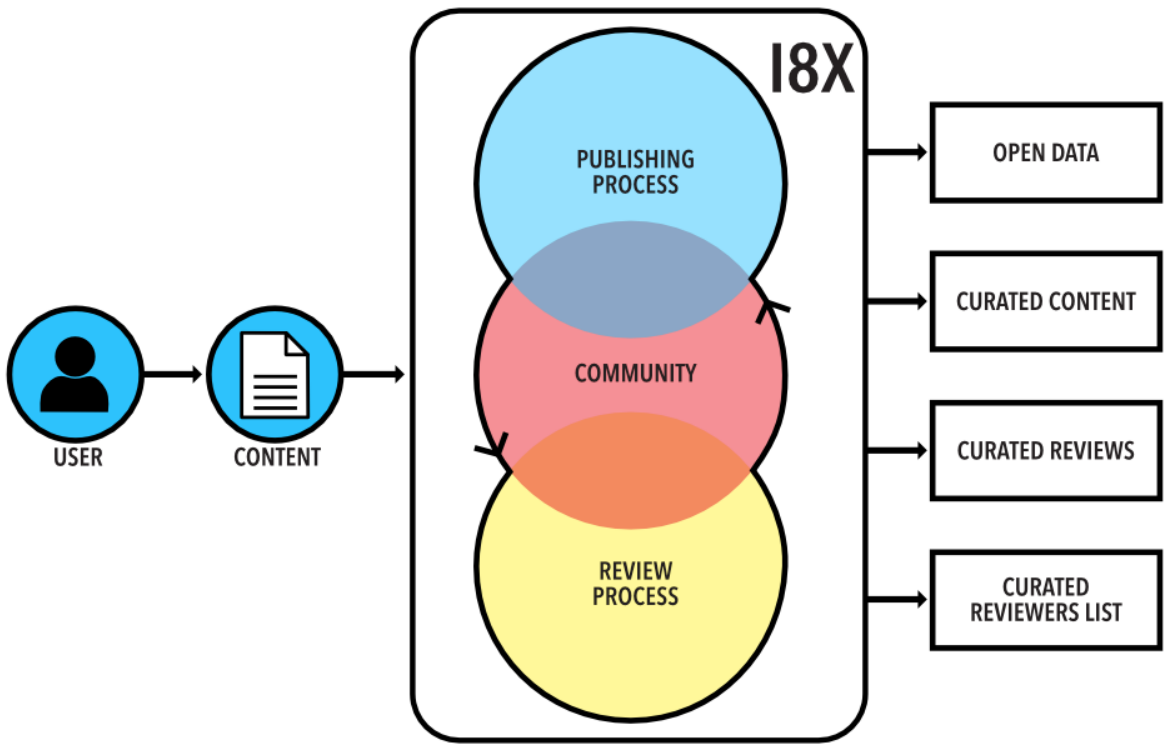

Figure 3: Functional overview of the I8X platform.

The main function of the I8X platform is to translate user generated content - scholarly communication (research paper, research idea, project proposal, ...) into curated and reviewed content. The output of the publishing and the review processes are curated reviews, curated list of trusted reviewers and open data access supported by I8X database of over 150 million metadata points on research papers and high quality index of keywords, authors, organisations and citations. The data on authors, reviews, reputation indices are stored and generated in a decentralized network and accessible to everyone through Open APIs or interaction with smart contracts provided and supported with our I8X technology stack.

There are three types of community members, each with specific roles in I8X platform:
- users,
- peers and
- developers.

Users are members of the community that interact with I8X platform through the application layer that provides WebUI access and functionality. User actions include claiming existing research articles and other published content, adding new content, reviewing and commenting on the content.

Peers are trusted members of the community that form and secure the blockchain network, execute transactions through consensus mechanisms and connect applications with the blockchain.

Developers are community members that use and/or build open APIs, data models and applications using I8X services. I8X applications are decentralized curated content collections. Some examples of applications are: a single research paper, decentralized journal, journal issue or book, decentralized conference proceedings, decentralized research project call.

One of the key parts of developing the I8X platform is the representation and encoding of content (paper, report, abstract, data, code) as smart contract on the blockchain. We take a two-step approach here: first we represent publishing phases of a research paper as a finite state machine, and then encode the process into a smart contract (Mavridou 2017).

Consider the process using publishing a research paper as an example. The process is started by the user announcing the intent to publish a paper by submitting the (possibly yet unfinished) content to the platform using one of the applications. The subset of content data (for instance authors, institutions, title, abstract) is hashed and written to the blockchain to secure authorship to authors. The review process (described below) is initiated by the author(s). The outcome of the review process can transition the paper into a published one or return the paper to author(s) for further revisions and development. Community can act also on already published content by raising an objective and propose the retraction of the published content.

To stimulate good behavior of users, to foster integrity and transparency of the network and to engage community members into actively following publishing and review processes, we integrated several socio-economic incentives and deterrents into the publishing and reviewing processes. As an example, users can use the I8X platform token for depositing in publishing and review processes. By successfully performing certain tasks on the platform such as publishing content, writing reviews and comments, users can earn tokens. The reputation of the user on I8X is a function of number of tokens the users hold at each instant of time. Since the number of tokens reflects the reputation of a user, they can exchange tokens only with I8X platform and not with other users.

A finite state machine (FSM) representing the paper has the following states (Figure 4): Active (A), Under review (U), Published (P), and Retracted (R).

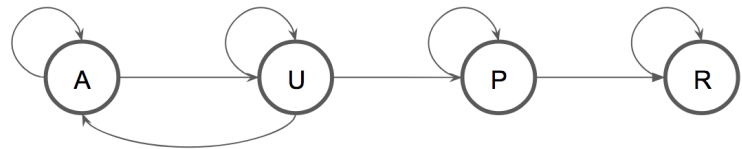

Figure 4: States and transitions in publishing process.

The FSM can change states with the transitions that are initiated by the community members. There are following transition in the publishing FSM:

A ==> A
In this state transitions are initiated by community comments which preserve the state. Community comments reflect expectations of the community about the research paper. There is no deposit required to submit comments in this state.

A ==> U
This transition is initiated by the author when the paper is completed and ready for review. The author deposits $N\_A > N\_A\_MIN$ tokens and this transaction triggers the start of the review process.

U ==> U

While the paper is in this state the review process is in progress. Community members can submit comments about the paper which preserve this state. Commenting in this state works like a prediction market (Thicke 2017; Arrow 2008; Hanson 1995; Hanson 2012) where users buy or sell shares for the outcome of the review process (revise or publish).

U ==> A

Initiated automatically by the platform as a result of reviewing process (with consensus by the listed reviewers), decision = revise. The author loses the deposited tokens, users that have bought revise shares receive reward.

U ==> P

Initiated automatically by the platform as a result of reviewing process (with consensus by the listed reviewers), decision = publish. The author is rewarded for successful publication by $N\_P > N\_A$ tokens, and users that have bought publish shares receive reward.

P ==> P

In this state users can submit community comments without token deposit and reviews through review process (see below) that can influence reputation of the reviewer and authors.

P ==> R

Usually an exceptional transition initiated by community member with "raise an objective" mechanism followed by a dispute/challenge. The outcome is decisioned by consensus of the peers (trusted community members).

R ===> R

R is a final state, users can submit comments without depositing tokens.

We base our infrastructure and protocol layers on the family of Hyperledger technologies licensed under the Apache 2.0 license and hosted by The Linux Foundation (https://www.hyperledger.org/). Using these technologies, recently proposed to be applied also in academic publishing field (Novotny, 2018), we are building a community blockchain infrastructure and protocols that will support the services of the platform and allow the development of applications.

As an example, we show here how a research paper FSM is modelled using the Hyperledger Composer framework. Hyperledger Composer supports writing transaction logic in JavaScript and asset modelling in Hyperledger Composer Modelling Language. The resulting application runs on Hyperledger Fabric infrastructure, a modular permissioned blockchain implementation also hosted by The Linux Foundation.

The code snippet below shows a simple modelling of an article asset, which consists of an article hash, its state, an array of owners (authors) and its Digital Object Identifier (DOI).

```
enum ArticleState {
      o ACTIVE
      o UNDER_REVIEW
      o PUBLISHED
      o RETRACTED
}

asset Article identified by articleHash {
      o String articleHash
      o ArticleState state
      --> User[] owners
      o String doi
}

transaction claimPublishedArticle {
      o String articleHash
      o String doi
}
```

The above code snippet also includes a transactions, which represents claiming ownership of an article already published on a different platform. The code snippet below shows transaction logic implementation as an asynchronous JavaScript function. The function first checks if the provided article already exists on the I8X distributed ledger and then updates the article registry accordingly.

```
async function claimPublishedArticle(tx) {
	let factory = getFactory();

	//determine if article already exists
	let articleRegistry = await
getAssetRegistry("infinitcodex.biznet.articles.Article");
	let exists = await articleRegistry.exists(tx.articleHash);

	let currentParticipant = getCurrentParticipant();
	if (exists) {
		//update the article
		let currentArticle = await articleRegistry.get(tx.articleHash);
		currentArticle.owners.forEach((owner) => {
			if (owner.$identifier === currentParticipant.userId){
				throw new Error("Owner has already claimed that article");
			}
		});

		currentArticle.owners.push(currentParticipant);
		await articleRegistry.update(currentArticle);
}
	else {
		//create new article
		let currentArticle = factory.newResource("infinitcodex.biznet.articles", "Article", tx.articleHash);
		currentArticle.doi = tx.doi;
		currentArticle.owners = [getCurrentParticipant()];
		currentArticle.state = "PUBLISHED";
		await articleRegistry.add(currentArticle);
	}
}
```

4. Conclusions

Blockchain is a disruptive solution for scholarly communication that "can potentially transform the current socio-technical stasis of moral and market crises associated with academic publishing" (Swist 2018). Current scientific publishing system relies on trusted third parties (journals, editorial boards, editors) that provide interactions between authors and reviewers, and there are conflicting incentives and lack of trust among participants. There is definitely a need for an objective, immutable history (or memory) of research results, and for a fair attribution and recognition of authorship that is verified through peer-to-peer consensus mechanisms.

We have presented our implementation of blockchain based platform for scholarly communication that stands on community driven publishing and reviewing processes and introduced research paper as a modifiable entity. Using a strategic game setting, we presented some of the social dilemmas occurring in academic publishing and showed that building the trustworthy scientific community is key for changing the current publish-or-perish culture into publish-and-flourish one.

Building a trusted peer-to-peer community requires transparency and right incentives for all participants. Various propositions for token-based economy and reward systems in decentralized scholarly publishing have already been advanced (Swan 2015), including incentives structure in peer review processes based on external cryptocurrency markets (Avital 2018). But what should proper incentives in scholarly communication really achieve? As we have argued here, we believe that whatever the mechanism, the goal must be to establish "a fair game for staking our reputation, so that on questions of interest to funders, we converge as fast as possible to the "right" answer." (Hanson 1995).

Chu, J.S. and Evans, J.A., 2018. Too Many Papers? Slowed Canonical Progress in Large Fields of Science.

Cooper, R., DeJong, D.V., Forsythe, R. and Ross, T.W., 1992. Communication in coordination games. *The Quarterly Journal of Economics*, *107*(2), pp.739-771.

Cooper, R., 1999. *Coordination games*. Cambridge University Press.

Crous, C.J., 2017. Could disruptive technologies also reform academia?. *Web Ecology*, *17*(2), pp.47-50.

Duh, A., Meznaric, S. and Korošak, D., 2016. Guerrilla media: Interactive social media. In *Media Convergence Handbook-Vol. 1* (pp. 307-324). Springer, Berlin, Heidelberg.

Duh, A., Slak Rupnik, M. and Korošak, D., 2018. Collective Behavior of Social Bots Is Encoded in Their Temporal Twitter Activity. *Big Data*, *6*(2), pp.113-123.

Economist 2015. The promise of the blockchain: The trust machine'. *The Economist.*, https://www.economist.com/leaders/2015/10/31/the-trust-machine

Edwards, M.A. and Roy, S., 2017. Academic research in the 21st century: Maintaining scientific integrity in a climate of perverse incentives and hypercompetition. *Environmental Engineering Science*, *34*(1), pp.51-61.

Ellison, G., 1994. Cooperation in the prisoner's dilemma with anonymous random matching. *The Review of Economic Studies*, *61*(3), pp.567-588.

Ellison, G., 2002. Evolving standards for academic publishing: A q-r theory. *Journal of political Economy*, *110*(5), pp.994-1034.

Ioannidis, J.P.A, Klavans, R., Boyack, K.W., 2018. Thousands of scientists publish a paper every five days. Nature 561, 167-169.

Janowicz, K., Regalia, B., Hitzler, P., Mai, G., Delbecque, S., Fröhlich, M., Martinent, P. and Lazarus, T., 2018. On the prospects of blockchain and distributed ledger technologies for open science and academic publishing. *Semantic Web*, (Preprint), pp.1-11.

Kirchherr, J., Why we can't trust academic journals to tell the scientific truth, 2017, The Guardian

Kiri, B., Lacetera, N. and Zirulia, L., 2018. Above a swamp: A theory of high-quality scientific production. *Research Policy*, *47*(5), pp.827-839.

Kuhn, S., 2017. Prisoner's Dilemma. *The Stanford Encyclopedia of Philosophy* (Spring 2017 Edition), Edward N. Zalta (ed.), https://plato.stanford.edu/archives/spr2017/entries/prisoner-dilemma/

Lacetera, N. and Zirulia, L., 2009. The economics of scientific misconduct. *The Journal of Law, Economics, & Organization*, *27*(3), pp.568-603.

Larivière, V., Haustein, S. and Mongeon, P., 2015. The oligopoly of academic publishers in the digital era. *PloS one*, *10*(6), p.e0127502.

Lazer, D.M., Baum, M.A., Benkler, Y., Berinsky, A.J., Greenhill, K.M., Menczer, F., Metzger, M.J., Nyhan, B., Pennycook, G., Rothschild, D. and Schudson, M., 2018. The science of fake news. *Science*, *359*(6380), pp.1094-1096.

Lee, C.J., Sugimoto, C.R., Zhang, G. and Cronin, B., 2013. Bias in peer review. *Journal of the American Society for Information Science and Technology*, *64*(1), pp.2-17.
23